\documentclass[aps,prl,twocolumn,amsmath,amssymb,showpacs,showkeys]{revtex4-1}
\usepackage{graphicx}
\usepackage{dcolumn}
\usepackage{bm}

\newcommand{\ra}{\right\rangle}        
\newcommand{\eq}[1]{Eq.~(\ref{#1})}    

\newcommand{\half}{\frac{1}{2}}

\begin{document}
\title{Quantum entanglement: The unitary 8-vertex braid matrix with imaginary rapidity}

\author{Amitabha Chakrabarti}
  \email{chakra@cpht.polytechnique.fr}
  \affiliation{Centre de Physique Th\'eorique, \'{E}cole Polytechnique, 91128 Palaiseau Cedex, France}
\author{Anirban Chakraborti}
  \email{anirban.chakraborti@ecp.fr}
  \affiliation{Laboratoire de Math\'{e}matiques Appliqu\'{e}es aux Syst\`{e}mes, \'{E}cole Centrale Paris, 92290 Ch\^{a}tenay-Malabry, France}
\author{Aymen Jedidi}
  \email{aymen.jedidi@ecp.fr}
  \affiliation{Laboratoire de Math\'{e}matiques Appliqu\'{e}es aux Syst\`{e}mes, \'{E}cole Centrale Paris, 92290 Ch\^{a}tenay-Malabry, France}

\date{\today}

\begin{abstract}
We study quantum entanglements induced on product states by the action of 8-vertex braid matrices, rendered unitary with purely imaginary spectral parameters (rapidity). The unitarity is displayed via the ``canonical factorization'' of the coefficients of the projectors spanning the basis. This adds one more \textit{new} facet to the famous and fascinating features of the 8-vertex model. The double periodicity and the analytic properties of the elliptic functions involved lead to a rich structure of the 3-tangle quantifying the entanglement. We thus explore the complex relationship between topological and quantum entanglement.

\end{abstract}



\maketitle

\setcounter{equation}{0}
``Are topological and quantum entanglements related?'' This intriguing question is recently being studied from different angles. One approach was intiated by Aravind \cite{Aravind}. Kauffman and Lomonaco \cite{Kauffman}  pointed out that braid matrices (representing the third Reidmeister move \cite{Reidemeister}, fundamental in the topological study of knots and links) correspond to universal quantum gates, when they are \emph{also} unitary. In previous studies unitary braid matrices were constructed explicitly for all dimensions \cite{Chakrabarti2007} and applied  to the study of quantum entanglements \cite{Chakrabarti2009}. Here our starting point is the 8-vertex model \cite{Baxter}  with braid matrix related to the Yang-Baxter one through a suitable permutation of elements, rendered unitary by a passage to imaginary rapidity  ($\theta \rightarrow i\theta$). The consequent unitarity is displayed transparently through ``canonical factorization'' \cite{Chakrabarti2003} of the coefficients of the projectors.  One now no longer has a statistical model with real, positive Boltzmann weights but unitarity thus implemented, opens a new road (as will be shown below) to quantum entanglements.
We first formulate such unitarization ($\theta \rightarrow i\theta$) in a general fashion and illustrate with the relatively simple 6-vertex case. Then we concentrate on the far more complex 8-vertex case, and study the 3-tangle \cite{Coffman} parametrized by sums of products of ratios of the $ q $-Pochhammer functions.

The $\widehat{R}\left(\theta\right)$ being a $N^2\times N^2$ matrix, acts on the base space $V_N\otimes V_N$ spanned by the tensor product of $ N $-dimensional vectors $ V_N $. Defining 
$
\widehat{R}_{12}\left(\theta\right)=\widehat{R}\left(\theta\right)\otimes I_N,
\widehat{R}_{23}\left(\theta\right)=I_N\otimes
\widehat{R}\left(\theta\right),
$
where $I_N$ is the $N\times N$ identity matrix, the corresponding braid operator is
\begin{eqnarray}
\widehat{B} & \equiv & \widehat{R}_{12}\left(\theta\right)\widehat{R}_{23}\left(\theta+\theta'\right)\widehat{R}_{12}\left(\theta'\right) \notag \\
&= &\widehat{R}_{23}\left(\theta'\right)\widehat{R}_{12}\left(\theta+\theta'\right)\widehat{R}_{23}\left(\theta\right).
\label{eq1.1}
\end{eqnarray}
The above Braid equation corresponds to the equivalence of knots related through the third Reidemeister move \cite{Reidemeister}.  Ref. \cite{yangbaxter} provides an useful introduction to the equivalent Yang-Baxter formalism. Of course, $\widehat{B}$ acts on the base space $V_N\otimes V_N\otimes V_N$.
Additionally, if the braid matrix $\widehat{R}$ is also \textit{unitary}, then it induces unitary transformations in $V_N\otimes V_N$, and $\widehat{B}$ in $V_N\otimes V_N\otimes V_N$. It is crucial to note the essential point that a non-trivial unitary $\widehat{R}$ induces \textit{non-local} unitary transformations. Had it been the case that
$\widehat{R}=\widehat{R_1}\otimes \widehat{R_2},$
where $\widehat{R_1}$ is acting on $ V_1 $, $\widehat{R_2}$ on $ V_2 $, and $\widehat{R}$ in
$ V_1 \otimes V_2 $, then such an $\widehat{R}$ would have been trivial from the point of braiding. Thus a non-trivial $\widehat{B}$ induces a non-local transformations in
$V_N\otimes V_N\otimes V_N$.

The non-local unitary actions set the stage for quantum entanglements.
It was shown \cite{Chakrabarti2009} that $\widehat{B}$, acting on \textit{un-entangled} product states of the general form
$$  \left|i\right\rangle \otimes \left|j\right\rangle \otimes \left|k\right\rangle \equiv \left(\sum_{i=1}^N
x_i\left|a_i\right\rangle\right)\otimes \left(\sum_{i=1}^N
y_i\left|b_i\right\rangle\right)\otimes\left(\sum_{i=1}^N
z_i\left|c_i\right\rangle\right)$$ in $V_N\otimes V_N\otimes V_N$, can generate entanglements for certain choices. We had also studied entanglements generated by two different classes (real and complex) of $\widehat{B}$. The ``3-tangles'' and ``2-tangles'' characterizing such entanglements were obtained explicitly in parametrized forms in terms of the parameters of $\widehat{B}$, and the variations with $(\theta,\theta')$ were analysed.

In another paper \cite{Chakrabarti2003}, we had introduced the ``canonical factorization'' for $\widehat{R} (\theta)$, which turns out to be very significant. 
Here we will exploit all these results and show that the simple passage $(\theta \rightarrow i \theta)$ is sufficient to provide unitarity under the following  constraints:
\begin{enumerate}
\item
$\widehat{R}\left(\theta\right)=\sum_{i}\dfrac{f_i\left(\theta\right)}{f_i\left(-\theta\right)}P_{i},$
where
$P_{i}P_{j}=\delta_{ij} P_{i},$
and
$\sum_{i} P_i =I_{N^2}\left(=I_N \otimes I_N \right).$
\item
$(\widehat{R}\left(\theta\right))_{\mathrm{trans}}=\widehat{R}\left(\theta\right).$
\end{enumerate}
Thus, initially $\widehat{R}\left(\theta\right)$ is \textit{real} and \textit{symmetric},
with a \textit{complete} set of orthonormal projectors $ P_i $ as a basis. The domain of $ i $
depends on the class considered.
The factorized form ${f_i\left(\theta\right)}/{f_i\left(-\theta\right)}$ of the coefficients in the first constraint
might seem strongly restrictive, but in fact it was shown that it holds true for all well-known standard cases, and new such cases were constructed \cite{Chakrabarti2003}, with the new term ``canonical factorization'' being introduced. One can easily check that a direct consequence of the constraints
is $\widehat{R}\left(\theta\right)\widehat{R}\left(-\theta\right) = I \otimes I.$ After the passage $(\theta \rightarrow i \theta)$, since $ P_j $'s are real, one can easily show that
$(\widehat{R}\left(i\theta\right))^\dagger (\widehat{R}\left(i\theta\right)) = I \otimes I$, i.e., $\widehat{R}\left(i\theta\right)$ is \textit{unitary}.

First, we demonstrate this formalism with the simpler case of the 6-vertex models. Following Ref. \cite{Chakrabarti2003}, which contains an extensive classification of ``canonical factorization'' for all dimensions, we define the projectors:
\begin{align}
P_{1 (\pm)}=\frac{1}{2}
\begin{vmatrix}
    1 & 0 & 0 & \pm 1 \\
    0 & 0 & 0 & 0 \\
    0 & 0 & 0 & 0 \\
    \pm 1 & 0 & 0 & 1
\end{vmatrix},
P_{2 (\pm)}=\frac{1}{2}
\begin{vmatrix}
    0 & 0 & 0 & 0 \\
    0 & 1 & \pm 1 & 0 \\
    0 & \pm 1 & 1 & 0 \\
    0 & 0 & 0 & 0
\end{vmatrix}
\label{eq9}
\end{align}
and obtain for the \textit{ferroelectric} case after $(\theta \rightarrow i \theta)$, with the real
parameter $ \gamma $,
\begin{align}
\widehat{R}\left(i\theta\right)= P_{1 (+)}+P_{1 (-)}+ \frac{\cosh{\frac{1}{2}(\gamma - i\theta)}}{\cosh{\frac{1}{2}(\gamma + i\theta)}} P_{2 (+)} \nonumber \\+ \frac{\sinh{\frac{1}{2}(\gamma - i\theta)}}{\sinh{\frac{1}{2}(\gamma + i\theta)}} P_{2 (-)},
\label{eq10}
\end{align}
which evidently satisfies the unitarity constraint.

Now, we proceed to the more complicated case of the 8-vertex model. We again define the projectors as in \eq{eq9}. The coefficients are expressed \cite{Jimbo} in terms of infinite products ($q$-Pochhammer functions), starting with
\begin{align}
\left(x;a\right)_{\infty}=\prod_{n\geq 0} \left(1-xa^n \right).
\label{eq11}
\end{align}
Setting $ z= \exp (\theta ) $, the initial 8-vertex matrix is:
\begin{align}
\widehat{R}\left(\theta\right)= \left( a+d \right) P_{1 (+)}+ \left( a-d \right) P_{1 (-)}+  \left( c+b \right) P_{2 (+)} \nonumber \\+  \left( c-b \right) P_{2 (-)},
\label{eq12}
\end{align}
where with supplementary real parameters $ p,q $ one obtains \cite{Chakrabarti2003}:
\begin{eqnarray}
   \left( a\pm d\right) & = &\frac{ \left(\mp p^{\half}q^{-1}z;p\right)_{\infty} \left(\mp p^{\half} q z^{-1};p\right)_{\infty} } { \left(\mp p^{\half}q^{-1}z^{-1};p\right)_{\infty} \left(\mp p^{\half}q z;p\right)_{\infty}}
   \label{eq13}
\end{eqnarray}
\begin{align}
   \left( c\pm b\right)  & = &\frac{\left( q^{\half}z^{-\half}\pm q^{-\half}z^{\half}\right)}{\left( q^{\half}z^{\half}\pm q^{-\half}z^{-\half}\right)} \frac{ \left(\mp p q^{-1}z;p\right)_{\infty}\left(\mp p q z^{-1};p\right)_{\infty} } { \left(\mp p q^{-1}z^{-1};p\right)_{\infty} \left(\mp pqz;p\right)_{\infty}}.
     \label{eq14}
\end{align}
We note that defining the numerators of the two equations \eq{eq13} and \eq{eq14} as $ f_{1 (\pm)} (z) $ and $ f_{2 (\pm)} (z) $ respectively, and using the fact that $ z= \exp (\theta ) $, we can express them as
$ \left( a\pm d\right)  = \frac{f_{1 (\pm)} (z) }{f_{1 (\pm)} (z^{-1}) }$ and
$ \left( c\pm b\right)   = \frac{f_{2 (\pm)} (z) }{f_{2 (\pm)} (z^{-1}) } , $
which implies that the essential property of the coefficients, ``canonical factorization'', is preserved.

After $(\theta \rightarrow i \theta)$ passage, we thus have
$\left( a\pm d\right) = \frac{f_{1 (\pm)} (e^{i \theta}) }{f_{1 (\pm)} (e^{-i \theta })},$
and $\left( c\pm b\right)  = \frac{f_{2 (\pm)} (e^{i \theta}) }{f_{2 (\pm)} (e^{-i \theta })}.$
Since the other parameters are real, we can interpret the coefficients as new \textit{phases}
$\left( a\pm d\right) = e^{i \Psi_{(\pm)} }$
and $\left( c\pm b\right)  = e^{i \Phi_{(\pm)} } ,$
where the \textit{phase factors} $ (\Psi_{(\pm)} , \Phi_{(\pm)} ) $ are complicated functions of $ (p,q, \theta )$. Note also that the coefficients under complex conjugation become
$ \left( a\pm d\right)^*  = \frac{f_{1 (\pm)} (e^{-i \theta })}{f_{1 (\pm)} (e^{i \theta })} =  \left( a\pm d\right)^{-1} $ and $ \left( c\pm b\right)^*  =  \frac{f_{2 (\pm)} (e^{-i \theta })}{f_{2 (\pm)} (e^{i \theta })}= \left( c\pm b\right)^{-1}. $
Since the projectors are real and symmetric, we again have the unitarity $(\widehat{R}\left(i\theta\right))^\dagger \widehat{R}\left(i\theta\right)) = I \otimes I$.
This opens the door of a new domain as a generator of quantum entanglements, as shown hereafter.

Consider the base space that is 8-dimensional and spanned by the states
$  \left| \epsilon_1 \ra \otimes \left| \epsilon_2 \ra \otimes \left| \epsilon_3 \ra
    \equiv \left |\epsilon_1 \epsilon_2  \epsilon_3  \ra, $
where $\epsilon_i = \pm; \; i=1,2,3.$
We will adopt a notation $(\left| + \ra, \left| - \ra) \rightarrow (\left| 1 \ra, \left| \bar{1} \ra)$ that generalizes smoothly to higher spins.
The braid operator is
\begin{equation}
    \widehat{B} = \widehat{B}^\dagger = ( \widehat{R}(i\theta) \otimes I_2) ( I_2 \otimes \widehat{R}(i\theta + i \theta')) (\widehat{R}(i\theta') \otimes I_2),
    \label{eq22}
\end{equation}
and the matrix
\begin{equation}
\widehat{R}(i \theta)=
\begin{vmatrix}
    a & 0 & 0 & d \\
    0 & c & b & 0 \\
    0 & b & c & 0 \\
    d & 0 & 0 & a
\end{vmatrix}
\label{eq24}
\end{equation}
where $(a \pm d) = e^{i\Psi_{(\pm)}(\theta)}, (c \pm b) = e^{i\Phi_{(\pm)}(\theta)},$
and
\begin{eqnarray}
   e^{i\Psi_{(\pm)}(\theta)} && = \frac{ \left(\mp p^{\half}q^{-1}e^{i\theta};p\right)_{\infty} \left(\mp p^{\half} q e^{-i\theta};p\right)_{\infty} } { \left(\mp p^{\half}q^{-1}e^{-i\theta};p\right)_{\infty} \left(\mp p^{\half}q e^{i\theta};p\right)_{\infty}} \notag\\
   e^{i\Phi_{(\pm)}(\theta)} && = \frac { q^{\half} e^{-i\frac{\theta}{2}} \pm q^{-\half} e^{i\frac{\theta}{2} }} { q^{\half} e^{i\frac{\theta}{2} } \pm q^{-\half} e^{-i\frac{\theta}{2} }} \times \notag\\
     && \frac{ \left(\mp p q^{-1}e^{i\theta};p\right)_{\infty} \left(\mp p q e^{-i\theta};p\right)_{\infty} } { \left(\mp p q^{-1}e^{-i\theta};p\right)_{\infty} \left(\mp pq e^{i\theta};p\right)_{\infty}}.
     \label{eq33}
\end{eqnarray}
One crucial fact is that $\widehat{R}$ has non-zero elements only on the diagonal and the anti-diagonal. This effectively splits the base space into two 4-dimensional subspaces closed under the action of $\widehat{B}$. They are spanned respectively by
$ V_{(e)} \equiv (\left| 111 \ra, \left| 1 \bar{1} \bar{1} \ra, \left| \bar{1} 1 \bar{1} \ra, \left| \bar{1} \bar{1} 1 \ra) $ and $ V_{(o)} \equiv  (\left| \bar{1} \bar{1} \bar{1}  \ra, \left| \bar{1} 11  \ra  \left| 1  \bar{1} 1 \ra, \left| 11 \bar{1} \ra), $
corresponding to even and odd numbers of indices with bar.
Moreover, for say
\begin{equation}
    \widehat{B} \left|111\ra
    = \alpha_1 \left|111\ra
    + \beta_1 \left|1\bar{1}\bar{1}\ra
    + \gamma_1 \left|\bar{1}1\bar{1}\ra
    + \delta_1 \left|\bar{1}\bar{1}1\ra,
    \label{eq26}
\end{equation}
one has
\begin{equation}
    \widehat{B} \left|\bar{1} \bar{1} \bar{1}\ra
    = \alpha_1 \left|\bar{1} \bar{1} \bar{1}\ra
    + \beta_1 \left|\bar{1}11 \ra
    + \gamma_1 \left|1\bar{1}1\ra
    + \delta_1 \left|11\bar{1}\ra,
    \label{eq27}
\end{equation}
with the \emph{same coefficients} $(\alpha_1, \beta_1, \gamma_1, \delta_1)$. More generally, the symmetry of \eqref{eq24} ensures for
\begin{equation}
    \widehat{B} \left|ijk\ra
    = c_1 \left|ijk\ra
    + c_2 \left|i\bar{j}\bar{k}\ra
    + c_3 \left|\bar{i}j\bar{k}\ra
    + c_4 \left|\bar{i}\bar{j}k\ra,
    \label{eq28}
\end{equation}
with $i,j,k=(1\text{ or } \bar{1})$, the direct consequence
\begin{equation}
    \widehat{B} \left|\bar{i} \bar{j} \bar{k}\ra
    = c_1 \left|\bar{i} \bar{j} \bar{k}\ra
    + c_2 \left|\bar{i} j k \ra
    + c_3 \left|i \bar{j} k\ra
    + c_4 \left|i j \bar{k}\ra.
    \label{eq29}
\end{equation}
The coefficients are conserved as above for $(i,j,k)\rightarrow(\bar{i},\bar{j},\bar{k}).$ Thus it is sufficient to evaluate the action of $\widehat{B}$ on the subspace $V_{(e)}$ or $V_{(o)}$.

To study the behavior of density matrcies and 3-tangles, we explicitly consider the action of $\widehat{B}$ on the product state $\left|1\ra \otimes \left|1\ra \otimes \left|1\ra \equiv \left|111\ra,$ in the subspace $V_{(e)}$,
given by \eqref{eq26}. Some straightforward algebra gives
\begin{eqnarray}
    \alpha_1 = f_{(+)} f{'}_{(+)} f{''}_{(+)} +  f_{(-)} f{'}_{(-)} g{''}_{(+)} \notag \\
    \beta_1 = g_{(+)} f{'}_{(+)} f{''}_{(-)} +  g_{(-)} f{'}_{(-)} g{''}_{(-)} \notag\\
    \gamma_1 = g_{(-)} f{'}_{(+)} f{''}_{(-)} +  g_{(+)} f{'}_{(-)} g{''}_{(-)} \notag\\
    \delta_1 = f_{(-)} f{'}_{(+)} f{''}_{(+)} +  f_{(+)} f{'}_{(-)} g{''}_{(+)},
    \label{eq31}
\end{eqnarray}
where we have used the \textit{phase factors} $\Psi_{(\pm)}$ and $\Phi_{(\pm)}$ to define
\begin{eqnarray}
    f_{(\pm)} & = & \frac{e^{i\Psi_{(+)}(\theta)} \pm e^{i\Psi_{(-)}(\theta)} }{2}\notag \\
    f'_{(\pm)} & = & \frac{e^{i\Psi_{(+)}(\theta')} \pm e^{i\Psi_{(-)}(\theta')} }{2}\notag \\
    f''_{(\pm)} & = & \frac{e^{i\Psi_{(+)}(\theta+\theta')} \pm e^{i\Psi_{(-)}(\theta+\theta')} }{2}\notag \\
    g_{(\pm)} & = & \frac{e^{i\Phi_{(+)}(\theta)} \pm e^{i\Phi_{(-)}(\theta)} }{2} \notag \\
    g'_{(\pm)} & = & \frac{e^{i\Phi_{(+)}(\theta')} \pm e^{i\Phi_{(-)}(\theta')} }{2} \notag \\
    g''_{(\pm)} & = & \frac{e^{i\Phi_{(+)}(\theta + \theta')} \pm e^{i\Phi_{(-)}(\theta + \theta')} }{2},
    \label{eq23}
\end{eqnarray}
such that $( f,f{'},f{''})_{(\pm)}$ correspond respectively to arguments $( \theta,\theta ',(\theta + \theta') )$ with analogous notations for $( g,g{'},g{''} )_{(\pm)}$.
Starting with \eqref{eq26} and tracing out the third index, one obtains the density matrix
\begin{equation}
\rho_{12} =
\begin{vmatrix}
    \alpha_1 \alpha^*_1 & 0 & 0 & \alpha_1 \delta^*_1 \\
    0 & \beta_1 \beta^*_1 & \beta_1 \gamma^*_1 & 0 \\
    0 & \beta^*_1 \gamma^*_1 & \gamma_1 \gamma^*_1 & 0 \\
    \alpha^*_1 \delta_1 & 0 & 0 & \delta_1 \delta^*_1
\end{vmatrix}.
\label{eq34}
\end{equation}
Defining
\begin{equation}
    \tilde{\rho}_{12} = \begin{vmatrix} 0 & -i \\ i & 0 \end{vmatrix}
    \otimes \begin{vmatrix} 0 & -i \\ i & 0 \end{vmatrix}
    \rho^*_{12}
    \otimes \begin{vmatrix} 0 & -i \\ i & 0 \end{vmatrix}
    \otimes \begin{vmatrix} 0 & -i \\ i & 0 \end{vmatrix},
    \label{eq35}
\end{equation}
one then obtains the matrix
\begin{equation}
\rho_{12} \tilde{\rho}_{12} = 2
\begin{vmatrix}
    \alpha_1 \alpha^*_1 \delta_1 \delta^*_1 & 0 & 0 & \alpha^2_1 \alpha^*_1 \delta^*_1 \\
    0 & \beta_1 \beta^*_1 \gamma_1 \gamma^*_1 & \beta^2_1 \beta^*_1 \gamma^*_1 & 0 \\
    0 & \gamma^2_1 \beta^*_1 \gamma^*_1 & \beta_1 \beta^*_1  \gamma_1 \gamma^*_1 & 0 \\
    \delta^2_1 \alpha^*_1 \delta_1 & 0 & 0 & \alpha_1 \alpha^*_1 \delta_1 \delta^*_1
\end{vmatrix}.
\label{eq36}
\end{equation}
The matrix $\left( \rho_{12} \tilde{\rho}_{12} \right)$ has the following eigenstates
\begin{equation}
    \left| \begin{matrix} \frac{\alpha_1}{\delta_1}\\0\\0\\1 \end{matrix} \ra,
    \left| \begin{matrix} \frac{\alpha_1}{\delta_1}\\0\\0\\-1 \end{matrix} \ra,
    \left| \begin{matrix} 0\\\frac{\beta_1}{\gamma_1}\\1\\0 \end{matrix} \ra,
    \left| \begin{matrix} 0\\\frac{\beta_1}{\gamma_1}\\-1\\0 \end{matrix} \ra,
    \label{eqEigenstate}
\end{equation}
with the eigenvalues $4 \alpha_1 \alpha^*_1 \delta^*_1 \delta^*_1, 0, 4 \beta_1 \beta^*_1 \gamma^*_1 \gamma^*_1, 0$, respectively.
Implementing the results of \cite{Coffman} (as in \cite{Chakrabarti2009}), the 3-tangle, invariant under permutations of the subsystems $(1, 2, 3)$, is obtained as
\begin{equation}
    \tau_{123} = 16 \left( \alpha_1 \alpha^*_1 \beta_1 \beta^*_1 \gamma^*_1 \gamma^*_1 \delta_1 \delta^*_1 \right)^{\half}.
    \label{eq41}
\end{equation}
Due to the unitarity of $\widehat{B}$ (after $\theta \rightarrow i \theta$) in \eqref{eq26}
$\alpha_1 \alpha^*_1 + \beta_1 \beta^*_1 +  \gamma^*_1 \gamma^*_1 + \delta_1 \delta^*_1 = 1$ and $ 0 \leq \tau_{123} \leq 1. $
As the parameters $(p,q,\theta,\theta')$ vary, the 3-tangle $\tau_{123}$ varies in the domain $[0,1]$.
The doubly periodic elliptic functions involved, expressed in terms of the $q$-Pochhammer functions as in the ratios \eqref{eq33}, demand painstaking computations involving rather involved algebra. This is indeed the real attraction of the unitarized 8-vertex case. 

\begin{figure*}
\includegraphics[width=0.2\linewidth]{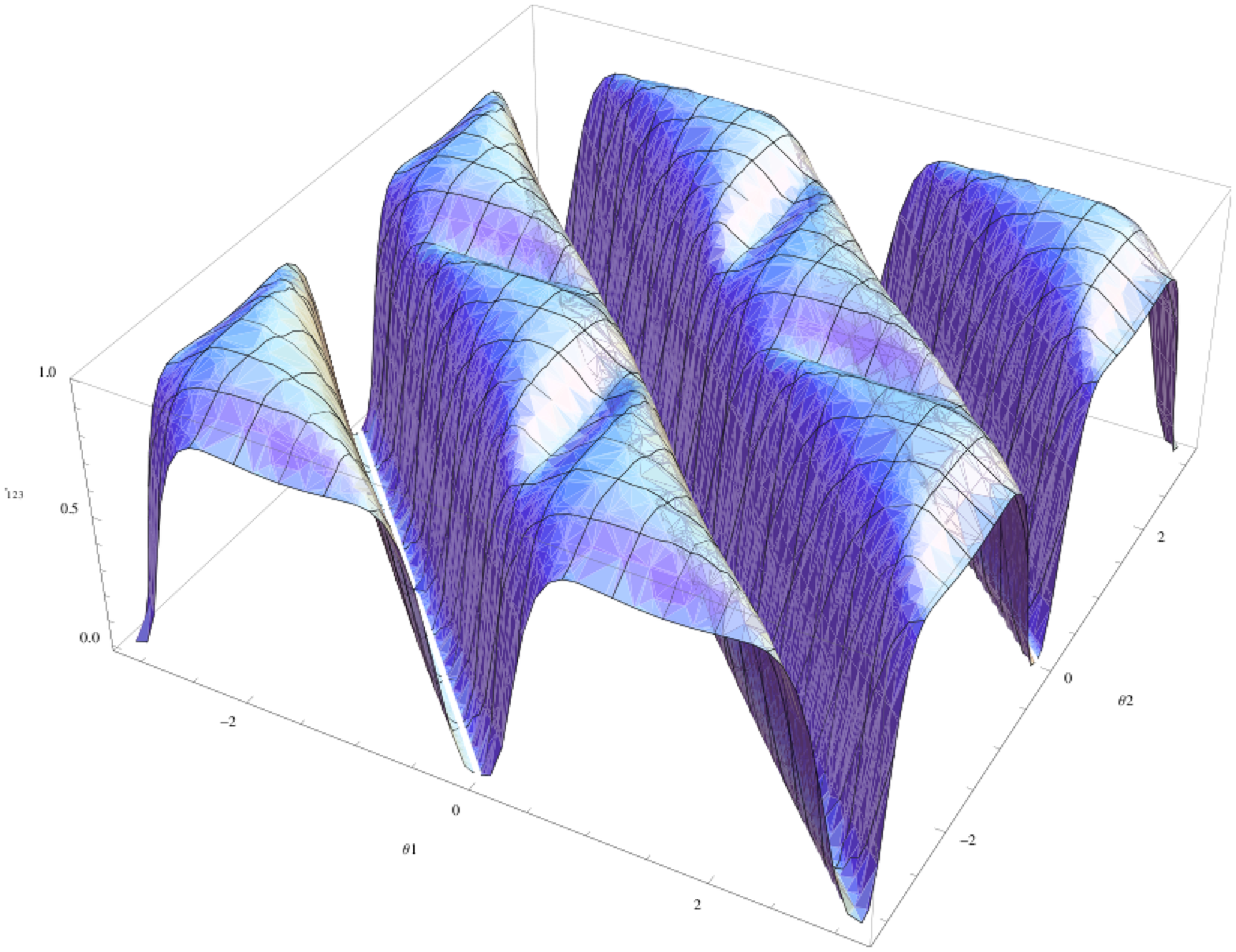}
\includegraphics[width=0.2\linewidth]{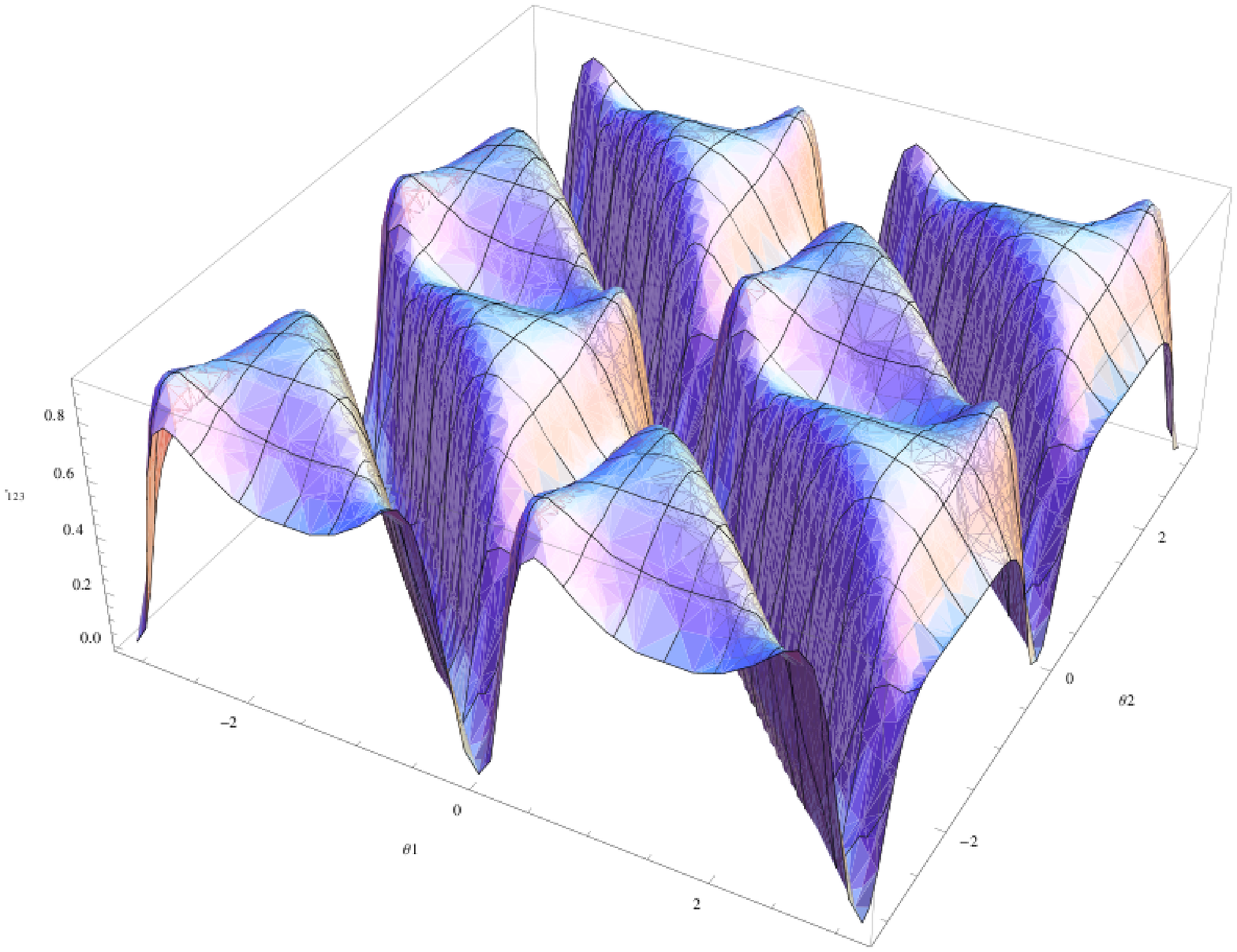}
\includegraphics[width=0.2\linewidth]{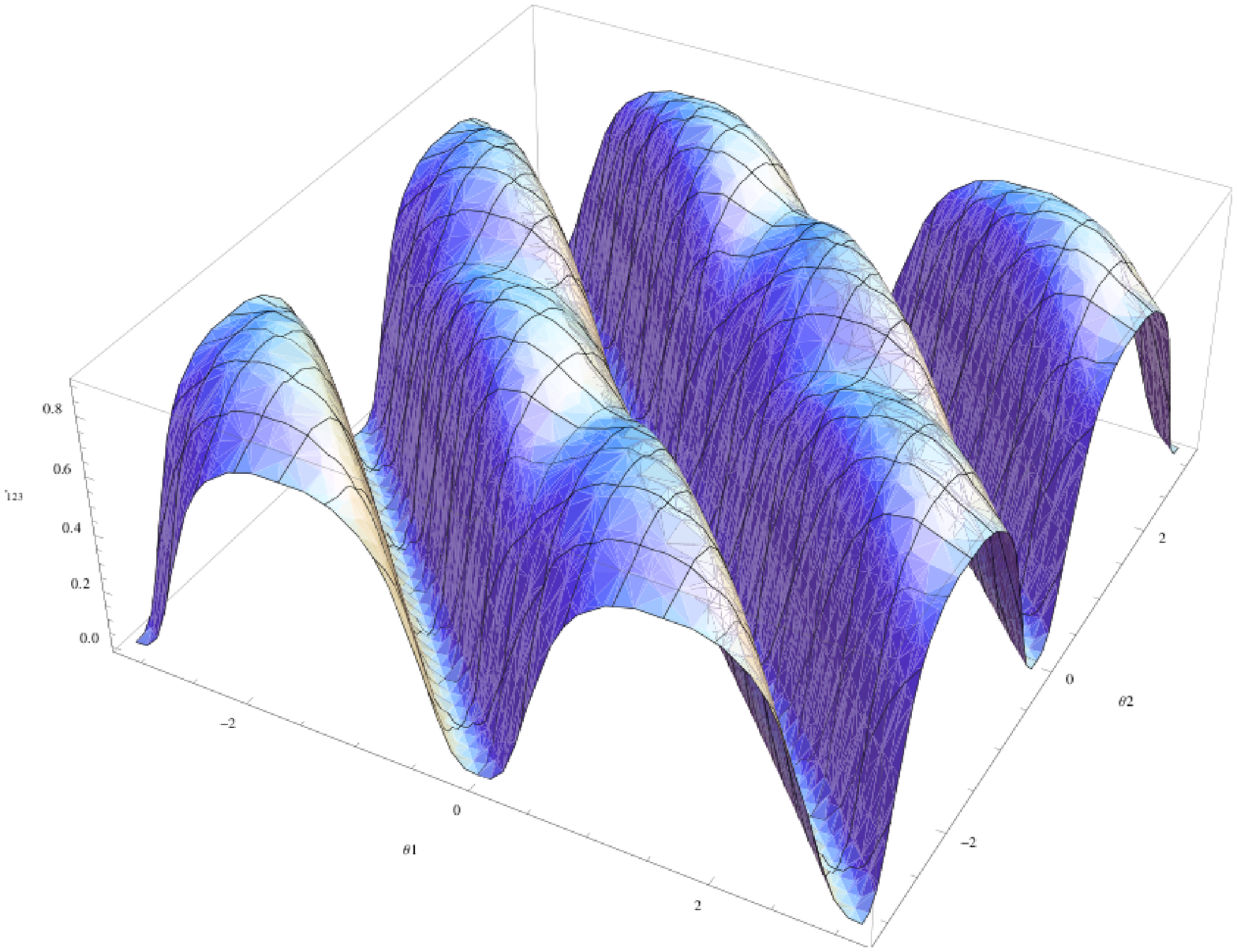}
\includegraphics[width=0.2\linewidth]{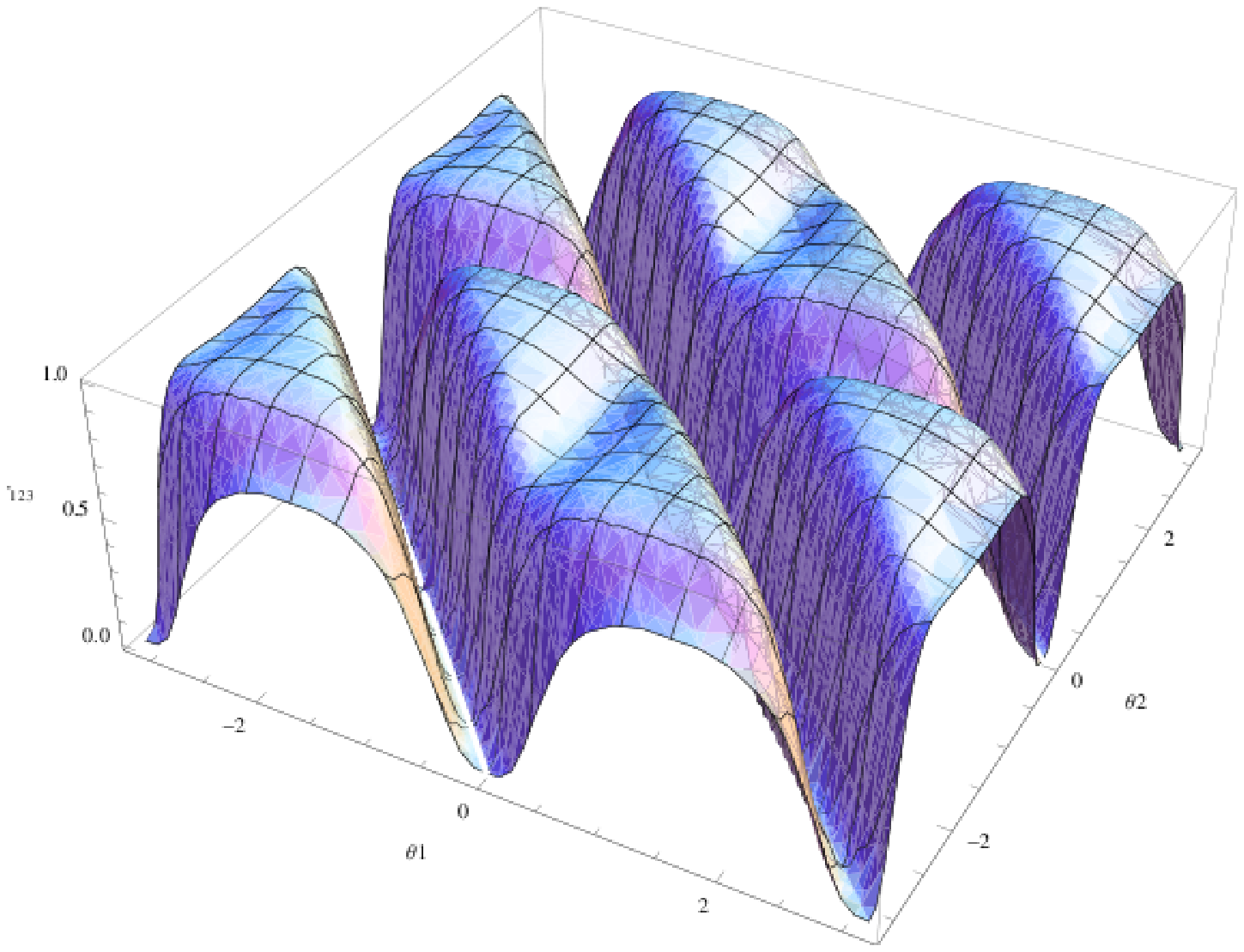}

\caption{\label{fig1}Variations of the 3-tangle $ \tau_{123} $ as a function of $(\theta, \theta')$, by the action of $ \widehat{B} $ on the product states $\left|111\ra, \left|1\bar{1}\bar{1}\ra, \left|\bar{1}1\bar{1}\ra, \left|\bar{1}\bar{1}1\ra$ in the subspace $V_{(e)}$, given by \eqref{eq26}. The parameters $ p=0.1, q=0.5 $.}
\end{figure*}


\begin{figure}
\includegraphics[width=0.75\linewidth]{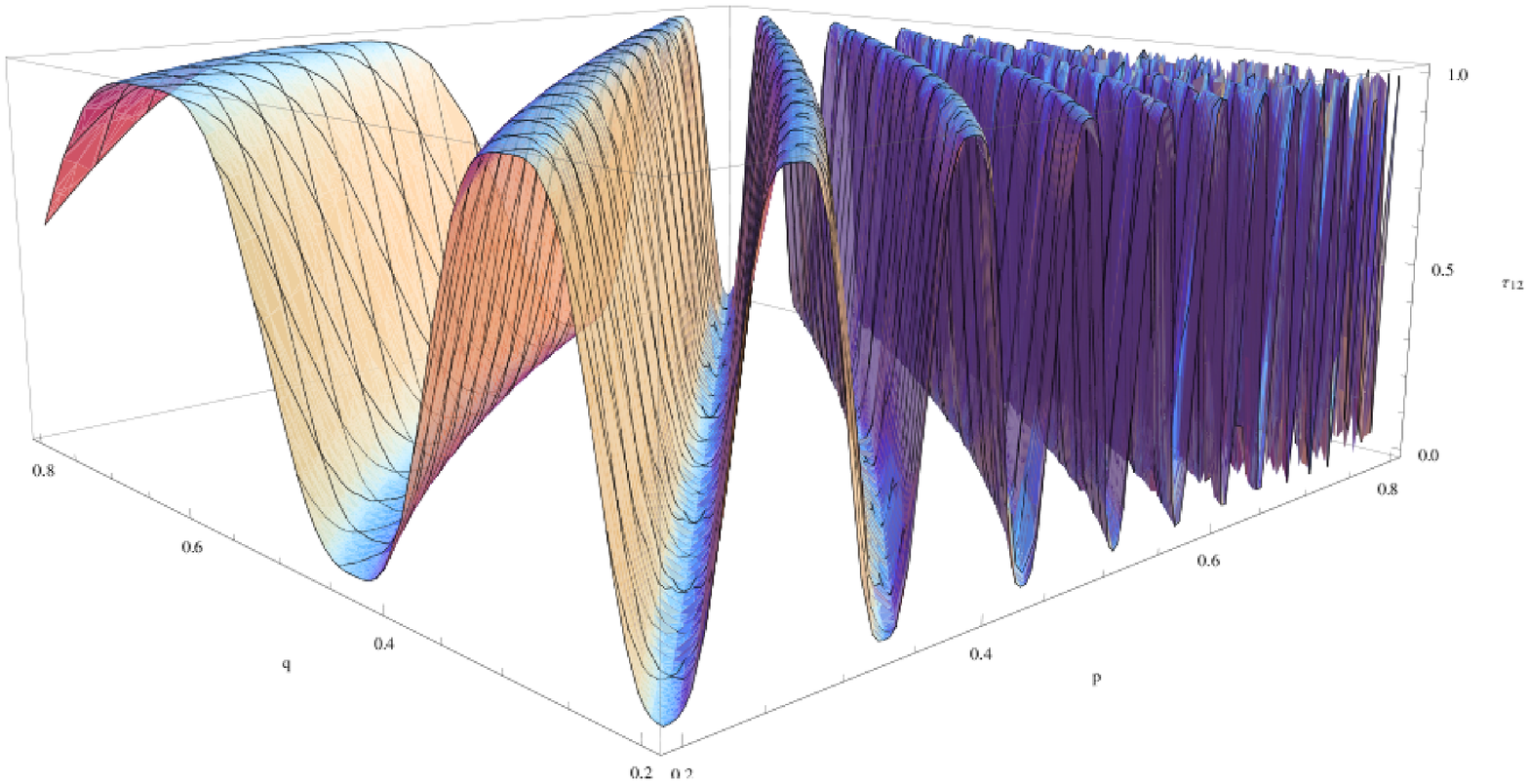}
\rotatebox{45}{\resizebox{!}{3.8cm}{%
   \includegraphics{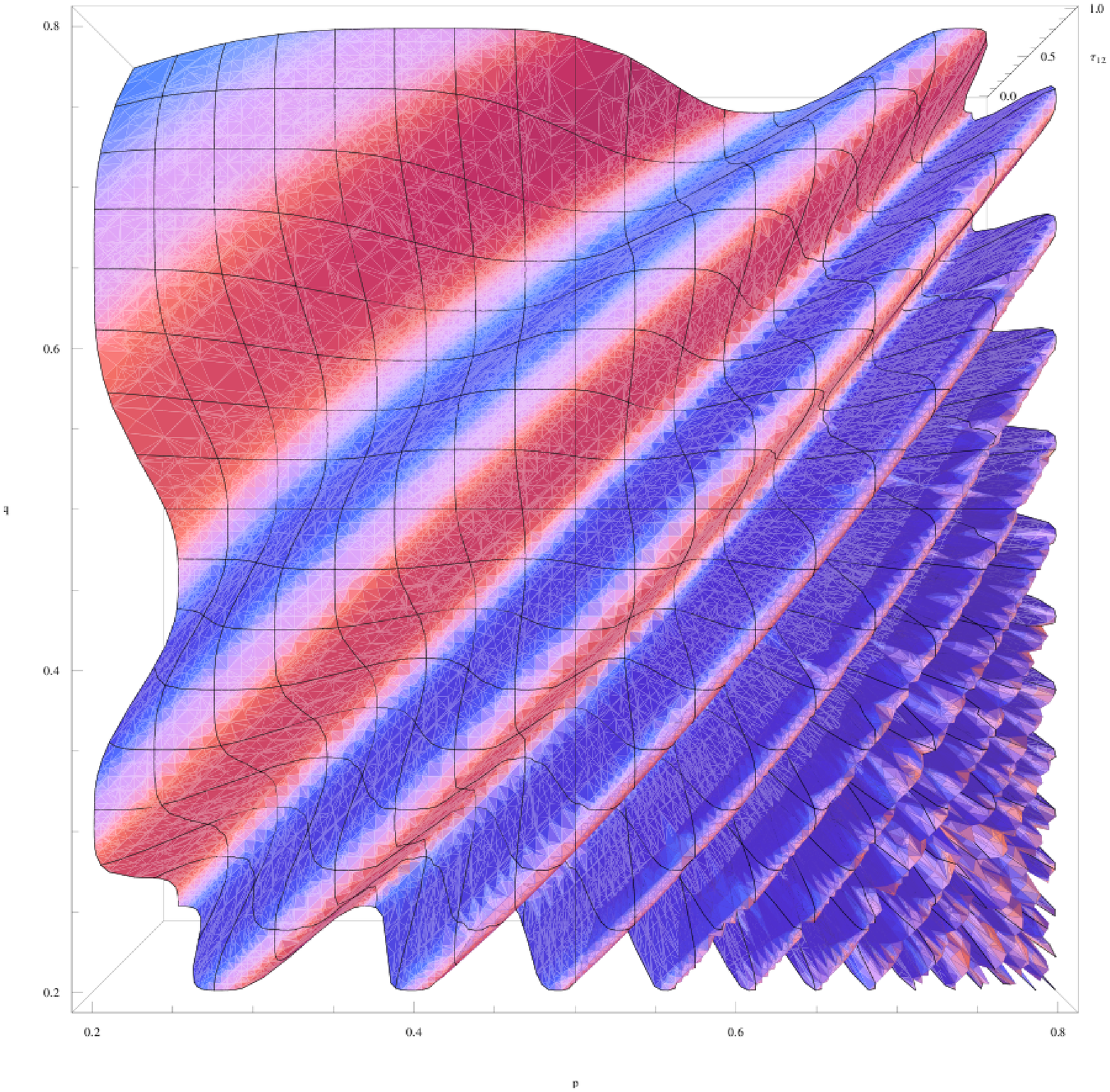}}}

\caption{\label{fig1}Variations (cross-sectional and top views) of the 3-tangle $ \tau_{123} $ as a function of $( p,q)$ for $(\theta=\pi /3, \theta'=\pi /6)$, by the action of $ \widehat{B} $ on the product state $\left|111\ra$.}
\end{figure}


One can study entirely analogously
$\widehat{B}( \left|1\bar{1}\bar{1}\ra, \left|\bar{1}1\bar{1}\ra, \left|\bar{1}\bar{1}1\ra )$ in the subspace $V_{(e)}$ implementing respectively the sets of coefficients
$(\alpha_i, \beta_i, \gamma_i, \delta_i), i=2,3,4$
as given by
\begin{eqnarray}
    \alpha_2 = f_{(+)} g{'}_{(+)} f{''}_{(-)} +  f_{(-)} g{'}_{(-)} g{''}_{(-)}, \notag \\
    \beta_2 = g_{(+)} g{'}_{(+)} f{''}_{(+)} +  g_{(-)} g{'}_{(-)} g{''}_{(+)} ,\notag\\
    \gamma_2 = g_{(-)} g{'}_{(+)} f{''}_{(+)} +  g_{(+)} g{'}_{(-)} g{''}_{(+)} ,\notag\\
    \delta_2 = f_{(-)} g{'}_{(+)} f{''}_{(+)} +  f_{(+)} g{'}_{(-)} g{''}_{(-)} ,\notag\\
    \alpha_3 = f_{(-)} g{'}_{(+)} g{''}_{(-)} +  f_{(+)} g{'}_{(-)} f{''}_{(-)} ,\notag \\
    \beta_3 = g_{(-)} g{'}_{(+)} g{''}_{(+)} +  g_{(+)} g{'}_{(-)} f{''}_{(+)} ,\notag\\
    \gamma_3 = g_{(+)} g{'}_{(+)} g{''}_{(+)} +  g_{(-)} g{'}_{(-)} f{''}_{(+)} ,\notag\\
    \delta_3 = f_{(+)} g{'}_{(+)} g{''}_{(-)} +  f_{(-)} g{'}_{(-)} f{''}_{(-)} ,\notag\\
    \alpha_4 = f_{(-)} f{'}_{(+)} g{''}_{(+)} +  f_{(+)} f{'}_{(-)} f{''}_{(+)} ,\notag \\
    \beta_4 = g_{(-)} f{'}_{(+)} g{''}_{(-)} +  g_{(+)} f{'}_{(-)} f{''}_{(-)} ,\notag\\
    \gamma_4 = g_{(+)} f{'}_{(+)} g{''}_{(-)} +  g_{(-)} f{'}_{(-)} f{''}_{(-)} ,\notag\\
    \delta_4 = f_{(+)} f{'}_{(+)} g{''}_{(+)} +  f_{(-)} f{'}_{(-)} f{''}_{(+)}.
\end{eqnarray}
Figure 1 shows the rich structure with subtle variations for $ \tau_{123} $, by the action of $ \widehat{B} $ on the product states $\left|111\ra, \left|1\bar{1}\bar{1}\ra, \left|\bar{1}1\bar{1}\ra, \left|\bar{1}\bar{1}1\ra$ in the subspace $V_{(e)}$. We note that for $(\theta + \theta' )= 0 $, we have $\tau_{123}= 0$, so that in the domain $(-\pi , \pi)$ for both  $(\theta, \theta')$, there are diagonal lines of  symmetry, with a line of zero value  passing  through the origin. One further notes that $(\Psi , \Phi )_{\pm} \rightarrow  ( \Psi , \Phi )_{\pm}$   and hence $ ( \tau_{123} \rightarrow \tau_{123} )$, for $( p \rightarrow 1/p , q \rightarrow 1/q, \theta \rightarrow - \theta )$. There are more intricate and subtle lines of symmetry as evident in figure 2, where we show the oscillations of $ \tau_{123} $ between zero and unity, as a function of $( p,q)$ for $ \widehat{B} \left|111\ra$.
The results for the other subspace $V_{(o)}$, namely
$\widehat{B}(\left| \bar{1} \bar{1} \bar{1}  \ra, \left| \bar{1} 11  \ra  \left| 1  \bar{1} 1 \ra, \left| 11 \bar{1} \ra),$
follows from the symmetry of $V_{(e)}$ and $V_{(o)}$ under the action of $\widehat{B}$ as stated in \eqref{eq28} and \eqref{eq29}. Combining these results one can then study the action of $\widehat{B}$ on the general product state, namely
$$\widehat{B}\left\{ (x_1 \left|1\ra + x_{\bar{1}} \left|\bar{1}\ra)
\otimes (y_1 \left|1\ra + y_{\bar{1}} \left|\bar{1}\ra)
\otimes (z_1 \left|1\ra + z_{\bar{1}} \left|\bar{1}\ra) \right\},$$
with some more straightforward algebra.


A central feature of quantum entanglements induced by unitary braid operators are parametrizations of the quantifiers of entanglements. The present case is a rich and subtle example. Various entangled cases chosen with simple constant coefficients to assure certain interesting properties (such as $ \left| GHZ  \ra$ and $ \left| W  \ra$) are thus seen to be imbedded in a continuum when approached via braid operators.  Such a continuum provides a link between topological and quantum entanglements. Some entanglements are inequivalent  under locally unitary transformations. Non-local unitary transformations, an intrinsic feature of braid matrices, provide precise explicit unifications. We intend to study all such aspects more thoroughly elsewhere.
Another quite different  perspective will be also shown to be provided by an entirely different class of braid matrices ($S$\textit{\^{O}}$(n)$-type \cite{Chakrabarti2005}) again ``unitarized'' by as above ($\theta \rightarrow i\theta$).  There will emerge a spin chain linked with  a class of  Temperley-Lieb algebra and display another possibility of our basic approach. Moreover, not being restricted to the ``diagonal-antidiagonal'' form (illustrated above by the 8-vertex model) these unitarized $S$\textit{\^{O}}$(n)$ matrices will generate a broader class of entanglements.


\end{document}